\documentclass[journal]{IEEEtran}
\usepackage{cite}
\usepackage{url}
\usepackage{longtable}
\usepackage{graphicx}
\usepackage{xfrac}
\usepackage{multirow}
\usepackage{amssymb}
\usepackage{float}
\usepackage{makecell}

\newcommand{\tabref}[1]{Table~\ref{#1}}
\newcommand{\eqnref}[1]{(\ref{#1})}

\newcommand{\figref}[1]{Figure~\ref{#1}}
\newcommand{\sfigref}[1]{Fig.~\ref{#1}}
\newcommand{\secref}[1]{Section~\ref{#1}}

\begin{document}

\title{Learning a Single Model with a Wide Range of Quality Factors for JPEG Image Artifacts Removal}

\author{Jianwei~Li,~\IEEEmembership{Member,~IEEE},
        Yongtao~Wang,~\IEEEmembership{Member,~IEEE},
        Haihua~Xie,
        and~Kai-Kuang~Ma,~\IEEEmembership{Fellow,~IEEE}
\thanks{This work was supported by National Key R\&D Program of China No. 2019YFB1406302.  This work was also a research achievement of Key Laboratory of Science, Technology and Standard in Press Industry (Key Laboratory of Intelligent Press Media Technology).
}
\thanks{Jianwei Li is with the Wangxuan Institute of Computer Technology, Peking University, Beijing 100080, China, and with the State Key Laboratory of Digital Publishing Technology, Peking University Founder Group Co., Ltd., Beijing 100871, China (e-mail: lijianwei@pku.edu.cn).
Yongtao Wang is with the Wangxuan Institute of Computer Technology, Peking University, Beijing 100080, China (e-mail: wyt@pku.edu.cn).
Haihua Xie is with the State Key Laboratory of Digital Publishing Technology, Peking University Founder Group Co., Ltd., Beijing 100871, China (e-mail: xiehh@founder.com).
Kai-Kuang Ma is with the School of Electrical and Electronic Engineering, Nanyang Technological University, Singapore 639798 (e-mail: ekkma@ntu.edu.sg).}
\thanks{Yongtao Wang is the corresponding author (e-mail: wyt@pku.edu.cn).}
}

\maketitle

\begin{abstract}
Lossy compression brings artifacts into the compressed image and degrades the visual quality. In recent years, many compression artifacts removal methods based on \emph{convolutional neural network} (CNN) have been developed with great success. However, these methods usually train a model based on \textit{one} specific value or a small range of quality factors. Obviously, if the test image¡¯s quality factor does not match to the assumed value range, then degraded performance will be resulted. With this motivation and further consideration of practical usage, a highly robust compression artifacts removal network is proposed in this paper. Our proposed network is a \emph{single model} approach that can be trained for handling a wide range of quality factors while consistently delivering superior or comparable image artifacts removal performance. To demonstrate, we focus on the JPEG compression with quality factors, ranging from 1 to 60. Note that a turnkey success of our proposed network lies in the novel utilization of the quantization tables as part of the training data. Furthermore, it has two branches in parallel---i.e., the \textit{restoration branch} and the \textit{global branch}. The former effectively removes the \textit{local} artifacts, such as ringing artifacts removal. On the other hand, the latter extracts the global features of the entire image that provides highly instrumental image quality improvement, especially effective on dealing with the \textit{global} artifacts, such as blocking, color shifting. Extensive experimental results performed on color and grayscale images have clearly demonstrated the effectiveness and efficacy of our proposed \textit{single}-model approach on the removal of compression artifacts from the decoded image.
\end{abstract}

\begin{IEEEkeywords}
compression artifacts removal, deep learning, quantization table.
\end{IEEEkeywords}

%
\IEEEpeerreviewmaketitle

\section{Introduction}

The goal of image compression or coding is to reduce the data size of the original digital image for the reduction of storage capacity and/or transmission bandwidth. This process incurs either no loss (i.e., \emph{lossless} coding) or acceptable loss (i.e., \emph{lossy} coding). Lossy compression algorithms, such as the well-known JPEG~\cite{JPEG-1992}, JPEG2000~\cite{jpeg2000}, WEB-P~\cite{WebP}, HEVC-MSP~\cite{HEVC-2012}, have been widely deployed in various types of electronic devices and systems. The encoded images could be required to further transmit the bitstream over the Internet. The amount of loss introduced in the lossy compression inevitably degrades the original image's quality and could produce annoying image artifacts in the decompressed or decoded image. All these degradations affect our viewing experience when such images displayed on a screen or printed on a paper. In fact, this could even affect the performance of other image processing tasks, such as image super-resolution~\cite{dong2015SRCNN}, character recognition~\cite{wang2012text}, to name a few. Therefore, how to remove image artifacts from the decoded images is of great importance to many image-based applications.

With the advent and fast advancement of \textit{convolutional neural networks} (CNNs), removal of image artifacts from the decoded images has been re-studied by exploiting this approach, and many state-of-the-art deep learning-based algorithms (e.g.,~\cite{AR-CNN-2015,multi-test2016,D3-2016,DDCN-2016,GAN-AR-2017,one-to-many-2017,cavigelli2017cas,zhang2018dmcnn}) have been developed with great success. Some algorithms achieve this objective only in the spatial domain~\cite{AR-CNN-2015,multi-test2016,GAN-AR-2017}, while others over the spatial and DCT domains~\cite{D3-2016,DDCN-2016}. For the former, a deep-learning neural network is exploited to learn a \emph{mapping function}, mapping from the decoded image to the original one (e.g., \cite{AR-CNN-2015}). For the latter, these methods are designed for the DCT-based compression algorithms (e.g., JPEG~\cite{JPEG-1992}), which is our main focus in this paper. The proposed methodology and developed techniques could be directly beneficial to other lossy compression algorithms and standards as well.

\begin{figure}[!t]
  \centering
  \includegraphics[width=1\linewidth]{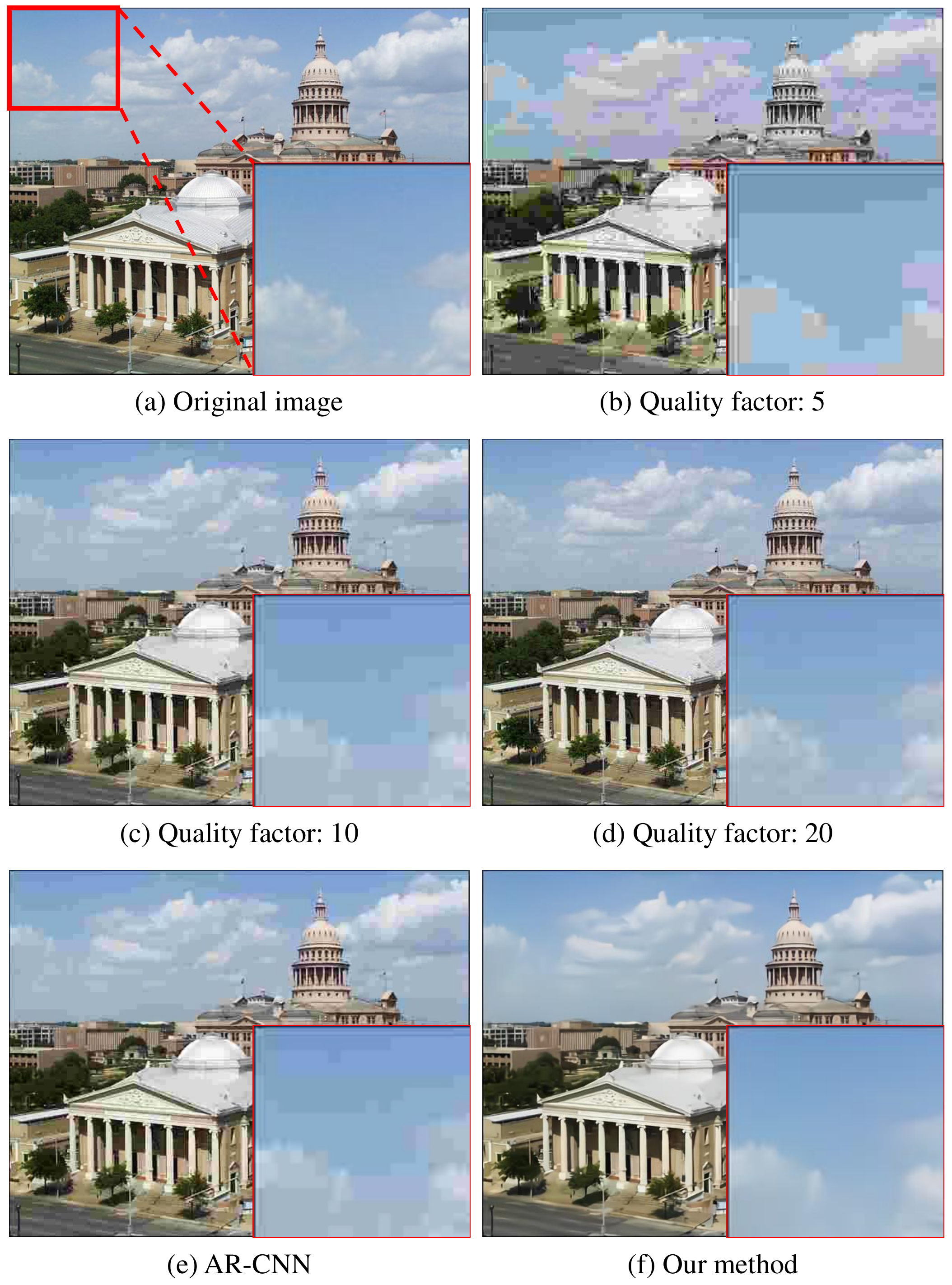}\\
  \caption{(a) An uncompressed color image. (b)-(d) A visual comparison of three JPEG-compressed images using different quality factors as indicated. (e)-(f) Two artifacts removal results for (c) (i.e., the quality factor 10) using the AR-CNN~\cite{AR-CNN-2015} and our method, respectively.}
  \label{fig:quality}
\end{figure}

In the JPEG lossy compression algorithm, users can decide the degree of compression by choosing a specific value of the \emph{quality factor} ranging from 1 to 100; the higher the value used, the better the image quality resulted (refer to~\figref{fig:quality}). It has been further noticed that most deep learning-based approaches establish their learning models for only \emph{one} specific value or a small range of quality factor~\cite{AR-CNN-2015,DDCN-2016,D3-2016,zheng2019IDCN}. As a result, the learned model can only effectively remove the resulted image artifacts from the decoded images compressed at the chosen quality factor, but yielding unsatisfied performance, in case other quality factors were used during the compression stage~\cite{multi-test2016}. Such mismatch should not be a surprise, since the resulted image degradations are quite different from each other under various quality factors imposed on the compression stage, as demonstrated in~\figref{fig:quality}.

Obviously, it would not be practical to design one learning model for \emph{each} quality factor individually. Hence, it would be ideal to design one \textit{single} model that can perform well for \textit{all} possible quality factors, from 1 to 100. To accomplish this challenging goal, two key novelties are introduced in this paper. First, the quantization tables are useful information as they directly reflect the image quality of the decoded image, and they are already available to access in the compressed image bitstream. This motivates us to utilize them as part of the inputs to train our network.

Second, the JPEG compression algorithm encodes images in terms of $8\times8$ blocks individually and independently. Therefore, the so-called \emph{blocking artifacts} tend to present in the decoded image when the compression is high. Furthermore, blocking artifacts tend to appear more distinctly on those image regions with smooth content. A set of such adjacently-clustered blocks could form the so-called \emph{layering artifacts} (e.g., the sky area in~\figref{fig:quality} (b)). To tackle this blocking artifacts problem, most of existing methods consider the use of larger-sized image patches (e.g., 32$\times$32 or 55$\times$55) for extracting more useful non-local features to reduce blocking artifacts more effectively~\cite{AR-CNN-2015,DDCN-2016, zhang2018dmcnn}.

However, the determined patch size might still not be large enough especially in the case of processing high-resolution images, let alone the above-mentioned layering artifacts could stretch from one end of the image to the other end. This motivates us to consider the \textit{entire} image on the extraction of its global features. For that, an additional neural network branch, denoted as the \emph{global branch}, is proposed in this paper and added in parallel with the main \emph{restoration branch}. The global branch is used to learn the global features from the entire decoded image. The learned knowledge is then exploited to assist the restoration branch to remove the compressed image artifacts more effectively.

The rest of the paper is organized as follows. Section~\ref{sec:related} reviews the literatures on image artifacts removal of the decoded images. Our proposed single model-based deep learning method is then introduced in Section~\ref{sec:method}. Extensive experimental results and comparisons with several state-of-the-arts are documented and discussed in Section~\ref{sec:exp} to demonstrate the efficacy and efficiency of our proposed method. Section~\ref{sec:conclude} concludes our paper.

\section{Related Works}
\label{sec:related}
Image restoration has been studied for a long time and many approaches have been developed~\cite{richardson1972bayesian, geman1987stochastic,li2015conformal,zhao2016loss,dong2018denoising,suganuma2019restore,zhang2019RNAN}. Compression artifacts removal is one of this kind of problems, aiming to reconstruct high-quality image from the decoded image. Traditional methods use digital filters to perform this task~\cite{reeve1984reduction,zakhor1992iterative,yang1993regularized,chen2001adaptive,zhang2013compression}. Reeve et al.~\cite{reeve1984reduction} reduces the blocking artifacts by applying a Gaussian filter to each block's boundaries. Chen et al.~\cite{chen2001adaptive} proposes an algorithm operated in the transform domain to alleviate the loss of transform coefficients to reduce blocking artifacts. Zhang et al.~\cite{zhang2013compression} reduces compression artifacts by estimating the transform coefficients of overlapped blocks from that of non-local blocks.

Learning-based approaches have been widely received and growing rapidly due to their impressive performance. Sparse representation is one of these techniques that can be exploited to remove image compression artifacts~\cite{jung2012sparse,chang2013dictionary,zhao2016reducing}. This kind of approach first learns a general dictionary of small patches from training images to represent the uncompressed image content. Then sparse coding algorithm is exploited to reconstruct the compressed image by the dictionary to reduce the compression artifacts. Liu et al.~\cite{liu2015data} proposes a sparse coding technique performed in the DCT and pixel domains jointly. They also use two prior information regarding the sparsity as well as the graph-signal smoothness to improve the restoration quality~\cite{liu2015inter}. However, the sparsity-based approaches are computationally intensive.

With the advent and fast developments of the CNNs, deep learning has been applied to many image restoration tasks successfully, including image compression artifacts removal~\cite{AR-CNN-2015,multi-test2016,DDCN-2016,D3-2016,GAN-AR-2017,one-to-many-2017,cavigelli2017cas,zhang2018dmcnn}. Dong et al.~\cite{AR-CNN-2015} first applies a CNN to perform image artifacts reduction using a four-layer end-to-end network. Svoboda et al.~\cite{multi-test2016} designs a deeper neural network with incorporation of residual learning and skip architecture. Guo et al.~\cite{DDCN-2016} designs a deep convolutional network working in the DCT and pixel domains. Such dual-domain presentation can make full use of the compression prior knowledge. Wang et al.~\cite{D3-2016} designs a deep dual-domain fast restoration model, which is inspired by the sparsity-based artifacts reduction method~\cite{liu2015data}. Galteri et al.~\cite{GAN-AR-2017} applies a generative adversarial network (GAN) in order to generate more pleasant results in visual quality. Guo et al.~\cite{one-to-many-2017} proposes a one-to-many network by selecting different loss functions.

Most of the deep learning based approaches train their networks according to a specific image quality level pre-assumed at the compression stage, hence they tend to yield poor performance once this assumption violates~\cite{multi-test2016,cavigelli2017cas}. In~\cite{multi-test2016,zheng2019IDCN}, the authors show a set of experimental results that are obtained from a CNN network trained from multiple degraded images with a small range of quality factors. However, the restoration results are inferior to the network trained with only one pre-determined quality level. Therefore, how to restore the decoded images suffered from various degrees of degradations by using one \textit{single} model is an open question and has been addressed in this paper as the main objective. Our proposed approach attempts to design a \textit{robust} deep learning network that will perform consistently well for all possible quality levels targeted at the compression stage. The \textit{robustness} here means that the designed \textit{single} model can perform equally well as that of other models that have been designed for only \textit{one} quality level in mind.

The idea of using a single model to handle multiple levels of degraded images has been developed in the work~\cite{zhang2018ffdnet}. The noise level map introduced in~\cite{zhang2018ffdnet} is similar to the quantization map generated by our proposed method. However, there are some differences as follows. First, the work~\cite{zhang2018ffdnet} is on image denoising for removing additive white Gaussian noise, while our method is on image artifacts removal for JPEG-compressed images. Although both are pursuing better image quality, however their problem¡¯s assumptions and the involved technical challenges are quite different. To our best knowledge, our method is the first work that is able to address a wide (even up to full) range of quality factors by using a single model, and yet consistently delivering impressive artifacts removal results with stable performance. Second, the work~\cite{zhang2018ffdnet} only assumes a fixed noise level, which is a single constant, to construct its noise level map. On the other hand, the quantization map generated in our method is constructed via two quantization tables (i.e., matrices), one for the luminance and the other for the chrominance, instead of the quality factors. Obviously, these matrices in our method have richer information than the use of a single constant in~\cite{zhang2018ffdnet}. Consequently, this will yield better network training. Last but not least, the noise level map in~\cite{zhang2018ffdnet} needs to be given by the users or estimated by using other algorithms before conducting its denoising task; clearly, this is not practical. In our method, the two quantization tables are already available in the JPEG bitstream, and therefore they are directly accessible from the JPEG image files.

\begin{figure*}[!t]
  \centering
  \includegraphics[width=1\linewidth]{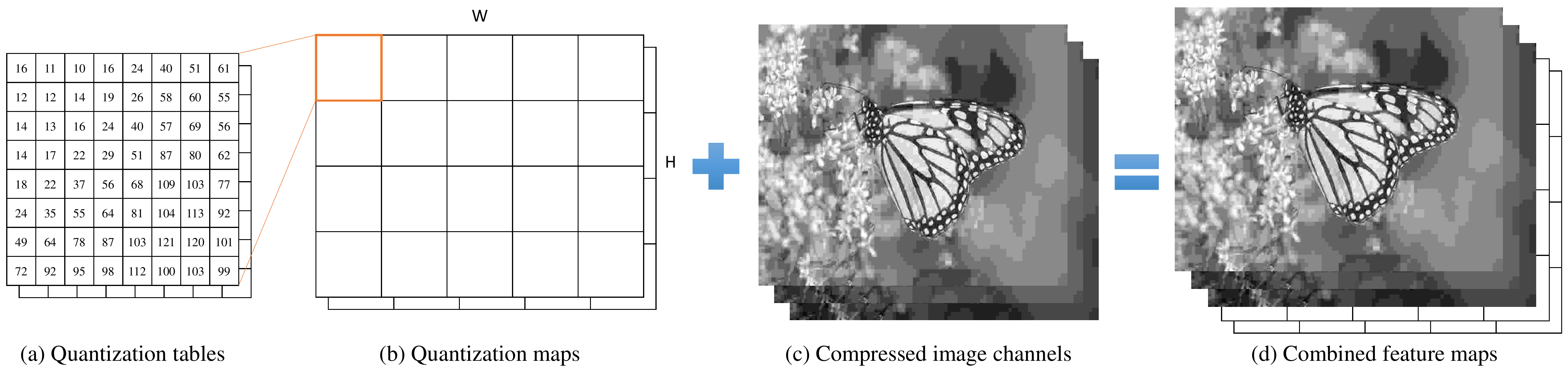}\\
  \caption{Incorporating the quantization tables into our network training. In JPEG-encoded color image, two \textit{quantization tables} (QTs) are used, one for the luminance (the $8\times8$ QT or matrix as shown in (a)) and the other for the chrominance (i.e., the one on the back of (a)). The \textit{quantization maps} (QMs) are constructed by tiling the QT through duplications as shown in (b), which has the same size as that of the compressed image as shown in (c). Combining the QMs in (b) and compressed image in (c), the \textit{feature maps}, as illustrated in (d), is formed and will be used for training our network.}
  \label{fig:quant}
\end{figure*}

\section{Proposed Method}
\label{sec:method}

\subsection{Motivation and Background}
Consider an original uncompressed image $\hat{X}$ with a dimension of $H\times W\times C$, where $H$ and $W$ denote the height and the width of the image, respectively, and $C$ is the number of image channels. For a typical monochrome image with 8-bits per pixel, the pixel values will be in the range of $[0, 255]$; likewise, a wider range $[0, 1023]$ for 10-bits medical images. In this paper, we shall focus on the former for ease and consistency of paper presentation. The compressed image $Y$ is generated by
\begin{equation}
\label{equ:compress}
  Y=F(\hat{X}; QF),
\end{equation}
where $F$ is the compression algorithm, $QF$ represents the \textit{quality factor} determined and used for adjusting the degree of compression. Now, the goal is to remove unwanted image artifacts that might appear on the compressed image $Y$. Hopefully, the restored image has a much-improved image quality, as close to $\hat{X}$ as possible; that is,
\begin{equation}
\label{equ:restore}
  X=G(Y)\approx\hat{X},
\end{equation}
where $G$ performs image artifacts removal function on $Y$ to reconstruct or restore a very high-quality image $X$ that is close to the ground-truth image $\hat{X}$. To tackle this problem, a convolutional neural network will be trained to learn a model for achieving the goal of image artifacts removal.

Given a training dataset with $N$ images, the loss function $L$ is defined below and to be minimized; that is,
\begin{equation}
\label{equ:lossfunc1}
  \hat{\theta}_g = \arg \min_{\theta } \frac{1}{N}\sum_{n=1}^{N} L\left ( G(Y;\theta_g),\hat{X} \right ),
\end{equation}
where $\theta_g$ represents the set of all parameters of the neural network. The loss function will be discussed in~\secref{sec:optimi} later with more details. In this paper, we shall focus on JPEG compression to demonstrate the efficacy and effectiveness of our proposed method on removing or reducing image artifacts introduced by the JPEG algorithm, especially when the degree of compression is high. Note that our method can be applied to other image compression algorithms that are similar to JPEG algorithm. In what follows, the \textit{lossy} JPEG compression algorithm~\cite{JPEG-1992} will be briefly reviewed.

The JPEG compression algorithm starts with dividing the original image into non-overlapping $8\times8$ blocks. Each image block is transformed to the frequency domain by applying the discrete cosine transform (DCT) to arrive at 64 DCT coefficients. The $8\times8$ DCT coefficients are then divided by an $8\times8$ \textit{quantization table} or matrix, followed by rounding the quotients to the nearest integers. After this quantization step, many DCT coefficient values become much smaller, and even to zeros. Consequently, this leads to much smaller number of non-zero DCT coefficients to be encoded for saving the storage. Since this is a lossy compression, the quality of the compressed image will be degraded; the larger the compression (i.e., using a smaller value of the $QF$ value), the more image distortion will be yielded. At last, the quantized DCT coefficients are further coded by using lossless coding algorithms (such as run-length encoding, Huffman coding, or arithmetic coding) and to form the final data bitstream according to the standardized bitstream syntax.

\subsection{Learning the Quantization Table for Network Training}
It has been noticed that most existing JPEG-compressed artifacts removal methods are developed based on \textit{one} specific or a small range of $QF$ values that was pre-determined and then used on the JPEG compression stage. Thus, these algorithms do not include $QF$ in \eqnref{equ:restore} and \eqnref{equ:lossfunc1} on their image restoration process. To equip our network with the ability on removing image artifacts for all the possibilities of $QF$ (i.e., from 1 to 100), the $QF$ is included in our network training from the outset to yield a generalized model that is able to deliver highly robust performance on removing image artifacts; thus, (\ref{equ:lossfunc1}) becomes
\begin{equation}
\label{equ:lossfunc2}
  \hat{\theta}_g = \arg \min_{\theta } \frac{1}{N}\sum_{n=1}^{N} L\left ( G(Y, QF; \theta_g),\hat{X} \right ).
\end{equation}
However, note that the $QF$ is just a user-defined \textit{scalar}-valued parameter, and therefore simply using the $QF$ alone is not effective to train a deep network. In our work, the user-determined $QF$ will be incorporated into \textit{quantization table} ($QT$) used in the JPEG compression. By imposing different values of $QF$, different $QT$s will be generated that will lead to different compressed image quality. The quantization table data is coded in the JPEG-compressed image data (i.e., bitstream), as it is required in the decoding process at the decoder to recover the pixel data from the binary bitstream. Thus, the $QT$ will be directly exploited in our network training process (refer to \figref{fig:quant}).

For encoding the color image, JPEG has two kinds of $QT$s, one for the luminance and the other for the chrominance. The $QT$ is a $8\times8$ matrix, which has the same size as that of an $8\times8$ block. For color image, the $QT$ has a dimension of $8\times8\times2$, where 2 is due to the luminance and chrominance. The $QT$ is further repeated in both the horizontal and the vertical directions to form a \textit{quantization map} ($QM$) with a dimension of $H\times W\times 2$, which has the same size as that of the original image (see \figref{fig:quant}(b)). Thus the quantization tables in $QM$ is aligned with the compression blocks in the compressed image\footnote{Note that if the image size is not a multiple integers of 8, the right-hand side and the bottom of the image will be padded with the pixel values duplicated from the boundaries of the image.}. The compressed image $Y$ is combined with $QM$ to form a new feature map $Y_{Q}$ with a dimension of $H\times W\times 5$. If the compressed image is grayscale, then both the image and $QM$ have only one channel each, and the dimension of $Y_Q$ is $H\times W\times 2$. Refer to \figref{fig:quant} for a graphical illustration. In our work, \eqnref{equ:lossfunc2} is changed to
\begin{equation}
\label{equ:lossfunc3}
  \hat{\theta}_g = \arg \min_{\theta } \frac{1}{N}\sum_{n=1}^{N} L\left ( G(Y_{Q};\theta_g),\hat{X} \right ).
\end{equation}
With this method, the neural network can learn the relationships between the quantization tables and different levels of artifacts of the compressed images. The learned model is able to handle any image artifacts removal for JPEG-compressed image with arbitrary $QF$ applied in the compression stage.

\begin{figure*}[!t]
  \centering
  \includegraphics[width=\textwidth]{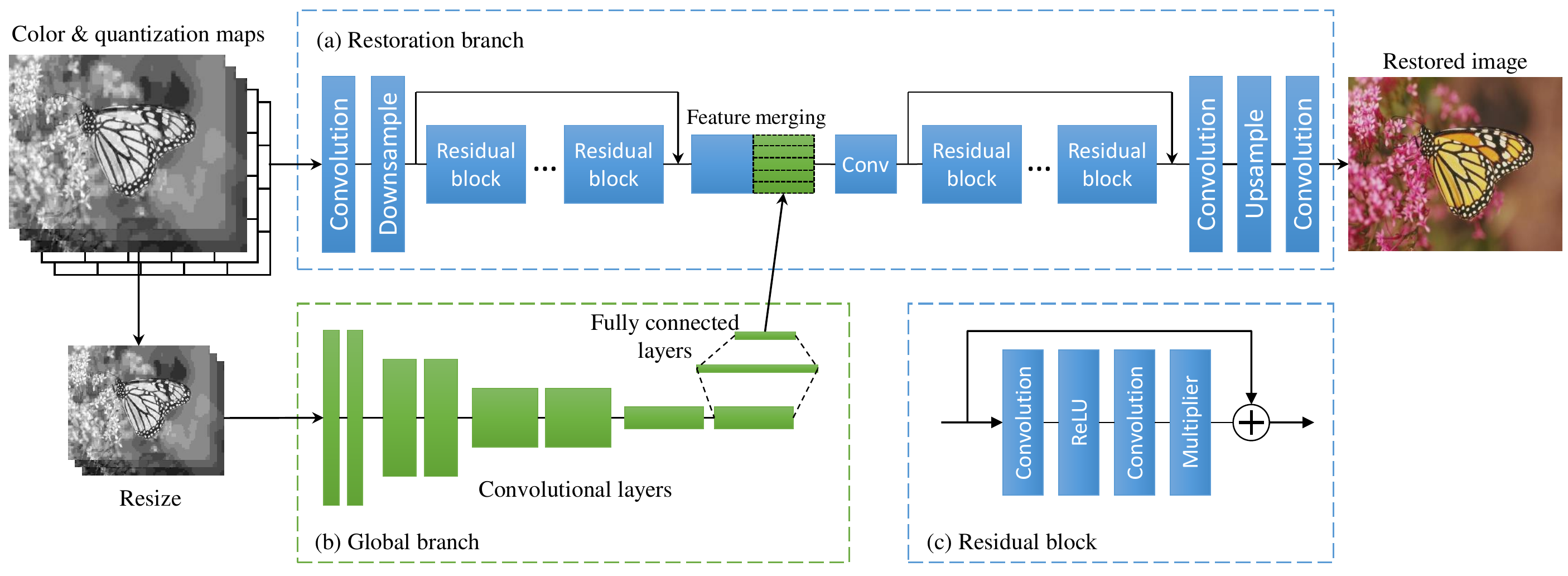}\\
  \caption{The network architecture of the proposed method. It contains the restoration branch (a) and the global branch (b). The restoration branch extracts local features and restores the compressed image. The global branch learns global features to improve the artifacts removal results. The global and local features are merged in the middle of the restoration branch. (c) is the structure of residual block in the network.}
  \label{fig:network}
\end{figure*}

\subsection{Proposed Network Architecture}
Our proposed network architecture consists of two branches: 1) the \textit{restoration} branch and 2) the \textit{global} branch, as shown in \figref{fig:network}. The restoration branch is the main part of the network that aims to remove image artifacts and improve image quality. It comprises several convolutional layers and residual blocks for extracting the \textit{local} features. In parallel with this part, the global branch proposed in this paper is used to extract the \textit{global} features, and these two types of features will be merged together in the middle of the restoration branch for further improving image artifacts removal results. More details are provided as follows.

\subsubsection{Restoration network branch}
Inspired by the single-image super-resolution method proposed in~\cite{EDSR-2017}, our restoration network branch is built upon the residual network~\cite{he2016residual}. The restoration branch includes several convolutional layers and residual blocks. Each residual block includes two convolutional layers and one ReLU activation layer, followed by a scaler layer in the end, as shown in \sfigref{fig:network}(c). The scaler layer is basically a multiplier, and the value of 0.1 has been empirically determined based on our simulation experiments. It is worthy to mention that the batch normalization layer~\cite{ioffe2015batch}, originally presented in the original residual network~\cite{he2016residual}, is able to improve the performance on high-level image tasks such as classification and recognition. However, it has been shown in~\cite{EDSR-2017} that, without exploiting the batch normalization layer, impressive performance on low-level image processing task can be achieved. For that reason, the batch normalization layer is removed from our residual block.

Our designed restoration branch network consists of two residual groups, where each residual group consists of a set of $N_{res}$ residual blocks together with a skip-connection. The first residual group is used to extract the local features from the compressed image, and the second residual group is exploited to process the local and global combined features in the middle of the restoration branch to complete the restoration process.

In order to reduce computation, the restoration branch first down-samples the image to one-quarter size of the image. After all processing, the feature maps need to be up-sampled back to the original size in the end of the network. The down-sampling process exploits a convolutional layer with a stride of two, and the up-sampling process adopts the sub-pixel convolution layer proposed in~\cite{shi2016sub-pixel}. The sub-pixel up-sampling method is able to avoid checkerboard artifacts which might be caused by the transposed convolutional layer.

In the restoration branch, the number of residual blocks $N_{res}$ is set to 32. The convolutional layers use the same settings; that is, the kernel size is set to $3\times3$, and the number of output channels is set to 64. As an exception, the last convolutional layer outputs the same channel as the input image (i.e., 3 for color image, 1 for grayscale image).

\begin{table}[!t]
\renewcommand{\arraystretch}{1}
\centering
\setlength{\abovecaptionskip}{0pt}
\caption{The settings of the global network branch.}
\label{tab:global}
\begin{tabular}{|c|c|c|c|c|c|}
  \hline
  Layer & Channel & Kernel & Stride & Padding \\ \hline
  conv1-1 & 32 & 4 & 2 & 1 \\ \hline
  conv1-2 & 32 & 3 & 1 & 1 \\ \hline
  conv2-1 & 64 & 4 & 2 & 1 \\ \hline
  conv2-2 & 64 & 3 & 1 & 1 \\ \hline
  conv3-1 & 128 & 4 & 2 & 1 \\ \hline
  conv3-2 & 128 & 3 & 1 & 1 \\ \hline
  conv4-1 & 256 & 4 & 2 & 1 \\ \hline
  conv4-2 & 256 & 3 & 1 & 1 \\ \hline
  fc1     & 1024 & - & - & - \\ \hline
  fc2     & 64 & - & - & - \\
  \hline
\end{tabular}
\end{table}

\subsubsection{Global network branch}
\label{sec:global}
The architecture of the global branch is shown in~\tabref{tab:global}, which consists of several convolutional layers at the beginning, followed by two fully-connected layers in the end. The structure of our global branch is motivated and evolved from image classification networks~\cite{alex2012alexnet,simonyan2014VGG}. The convolutional layers aim to extract features of the image, and the fully-connected layers output a vector, which represents the entire image, to fulfill the classification. In our global branch, the final linear layer will output a 64-dimensional vector to represent the global features of the compressed image. With consideration of fully-connected layers in our designed global network, the input image will be re-sized to the fixed resolution of $112\times112$ to avoid high computation complexity.

The extracted 64-dimensional global feature vector is then combined with the local feature maps extracted by the first group of residual blocks in the restoration network branch. The local feature maps are of $\sfrac{H}{2} \times \sfrac{W}{2} \times 64$ dimension. To merge the two features with different dimensions, the global feature vector will be repeated with $\sfrac{H}{2} \times \sfrac{W}{2}$ times, followed by stacking them into a 3-D volume with the same dimension as that of the local feature maps. Now, the extended global features and local features can be easily concatenated as new feature maps with a dimension of $\sfrac{H}{2} \times \sfrac{W}{2} \times 128$. The combined feature maps will be fed into the second group of residual blocks to conduct the final restoration process.

\subsection{Optimization}
\label{sec:optimi}
The loss function $L$ in \eqnref{equ:lossfunc3} is defined as
\begin{equation}
\label{equ:loss}
  L=\frac{1}{n} \sum_{i=1}^{n} \left \| X_i-\hat{X}_i \right \|_1,
\end{equation}
where $n$ is the number of the pixels, $\|\cdot\|_1$ denotes the $\ell_1$ norm. The reason of using the $\ell_1$ norm, instead of the mean-squared error, is because it can yield a sharper image result, as suggested in the open literatures (e.g.,~\cite{EDSR-2017}).

The global branch in our network learns features from the entire image. This means that a higher-resolution image with more details is beneficial to network training and achieves an improved performance, but on the expense of increasing computational complexity. To address this issue, the ``easy-hard transfer'' method as proposed in~\cite{AR-CNN-2015} is exploited to train our network gradually. In our work, the transfer learning is exploited on training patch size; that is, learning from small image patches and then transfer it to learning from large image patches. Specifically, the network is first trained on small image patches with a size of $64\times64$ to learn an initial model sufficiently. Then the image patches are enlarged to $256\times256$ to continue the training with the weights initialized by the above initial model. The former process trained on small patches mainly trains the restoration network branch to achieve the abilities of local features learning and artifacts removal. The latter process trained on large patches makes the global branch gain the ability to learn useful global features from the compressed images to improve the performance of image artifacts removal.

\section{Experiments}
\label{sec:exp}
\subsection{Simulation Experiment Setup}
\label{sec:implement}
The DIV2K dataset~\cite{DIV2K-2017} is exploited to train our network. This is a high-resolution image dataset established in workshop challenges of \cite{CVPR_Workshops_2017} and \cite{CVPR_Workshops_2018}. This dataset has been used for evaluating image denoising and super-resolution, and we consider it is also suitable for compression image artifacts removal. Specifically, we use its training set (800 images) for training, and validation set (100 images) for conducting performance evaluation of developed algorithm. Besides the DIV2K, another image dataset LIVE1 (29 images)~\cite{live1-2005,sheikh2006statistical} is also exploited as the test dataset to evaluate our algorithm's performance.

A variety of software packages offer image compression function. However, there is no standard criterion to generate the quantization table. In fact, different software may have their own default quantization tables. For example, Adobe has its own quantization tables in its products, and the algorithm is not disclosed to the public. In this paper, Python Image Library (PIL) is adopted to encode images into JPEG format for conducting all simulation experiments. PIL uses a standard quantization table proposed by the Independent JPEG Group~\cite{JPEG-Group}, which is adopted by most compression algorithms. It has a quality control factor, called the $QF$, ranging from 1 to 100, to adjust the quantization table. Our proposed CNN model is able to be trained for handling any wide range of compression quality factors, even including the full range, from 1 to 100, which covers all the possibilities of the JPEG compression. However, when QF $>$ 60, the compressed image is already in good/high quality. As a result, the decoded image is hard to show any concerned image artifacts for removal. Therefore, we focused on the JPEG-compressed images with $QF\in[1, 60]$ to evaluate the effectiveness of artifact removal in this paper.

\begin{figure*}[!ht]
  \centering
  \includegraphics[width=1\textwidth]{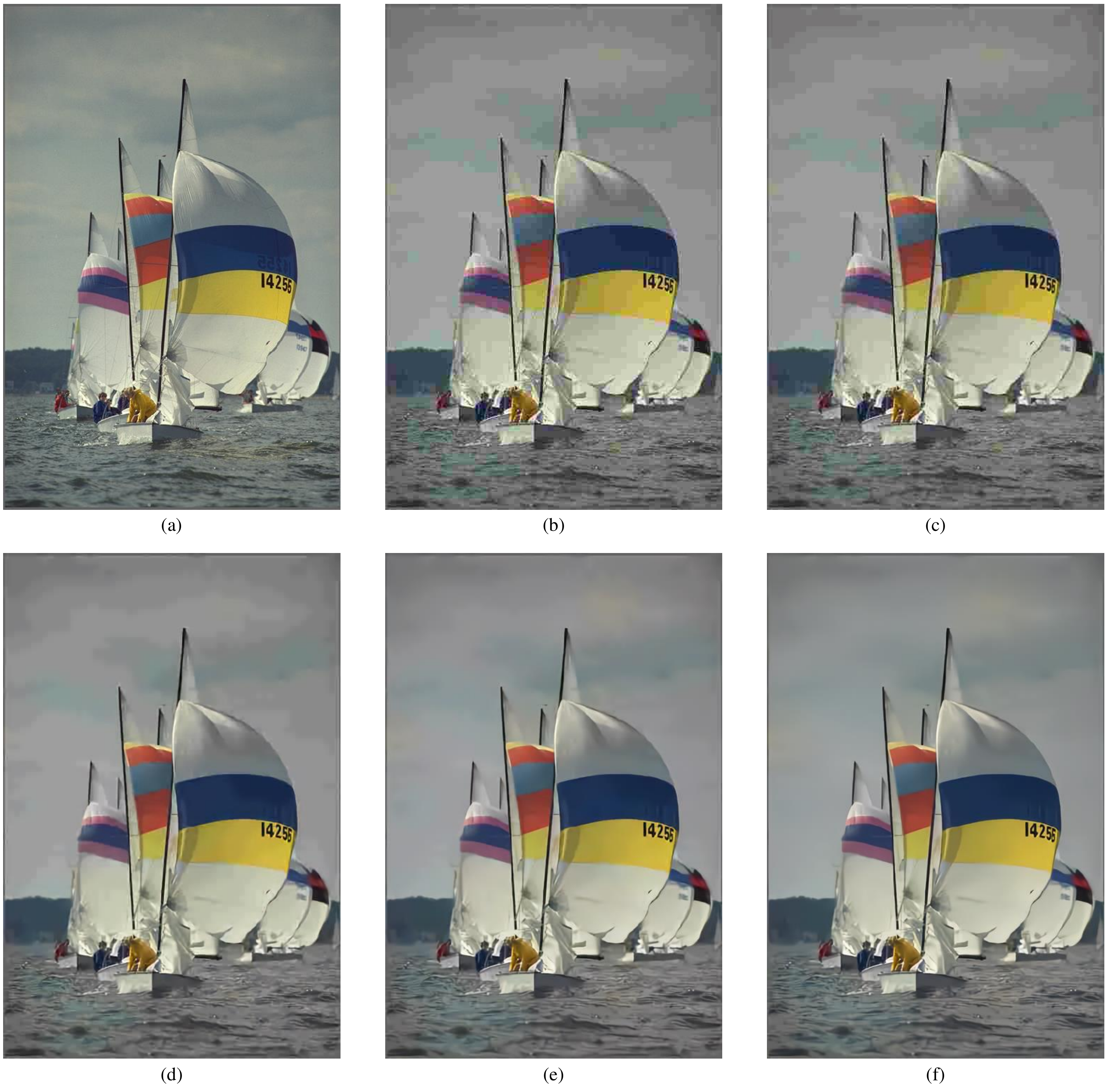}\\
  \caption{A visual comparison of image artifacts removal results using different methods, with an emphasis to highlight that our method is able to handle \textit{global} structure well. (a) the original image \textit{Sailing}, (b) the JPEG-encoded image in 28.28 dB, with the quality factor of 10 imposed on the JPEG compression stage, (c) AR-CNN method~\cite{AR-CNN-2015} in 29.02 dB, (d) EDSR method~\cite{EDSR-2017} in 30.38 dB, (e) our QCN method in 30.62 dB, and (f) our QGCN method in 31.01 dB. Note that the annoying contours presented in the sky area in (b) almost completely disappear in (f). Also the overall color image quality presented in (f) is quite near to that of the ground truth in (a). This example demonstrates the effectiveness and advantage of introducing the proposed global network branch in our method.
  }
  \label{fig:compare1}
\end{figure*}

\begin{figure*}[!ht]
  \centering
  \includegraphics[width=1\textwidth]{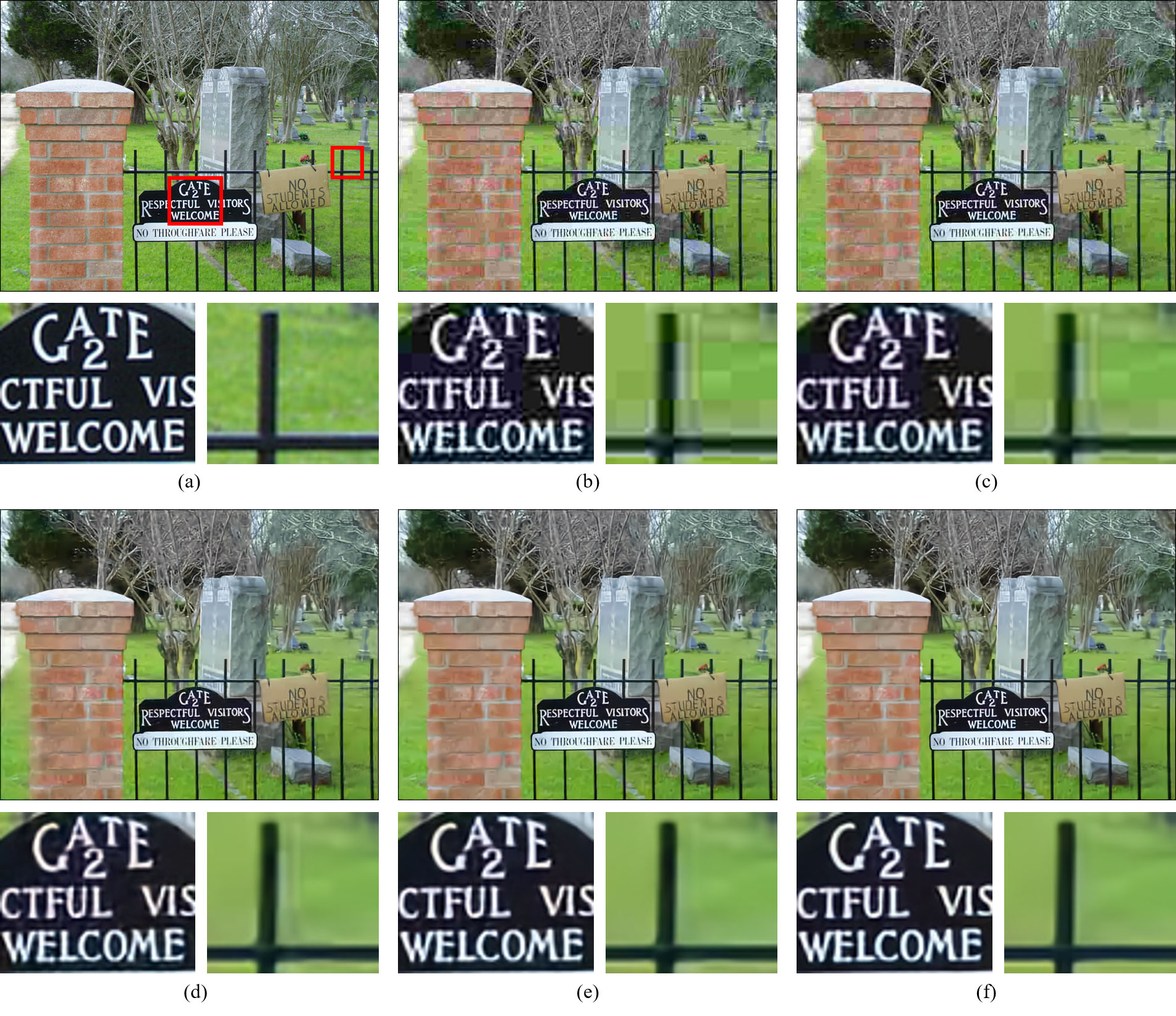}\\
  \caption{A visual comparison of image artifacts removal results using different methods, with an emphasis to highlight that our method is able to handle \textit{local} structure quite well. (a) the original image \textit{Cemetry}, (b) the JPEG-encoded image in 23.79 dB, with the quality factor of 10 imposed on the JPEG compression stage, (c) AR-CNN method~\cite{AR-CNN-2015} in 24.50 dB, (d) EDSR method~\cite{EDSR-2017} in 25.45 dB, (e) our QCN method in 25.64 dB, and (f) our QGCN method in 25.94 dB. Note that our QGCN method effectively removes the ringing artifacts (around the characters and gate) and blocking artifacts (on the grass background).}
  \label{fig:compare2}
\end{figure*}

Most existing compression artifacts removal methods train and test their models on the luminance component of color image (i.e., the Y channel of the YCbCr color space). This is because the human visual system is more sensitive to luminance than to that of chrominance. However, the chrominance channels are also subject to compression via the chrominance quantization table. Restoration of the chrominance channels can help improve the image quality. Our proposed method will put more focus on the evaluation of color images to show the effectiveness of our developed image artifacts removal algorithm. Experiments conducted on the grayscale images will be reported to compare with the existing methods.

Data augmentation is applied to the training dataset, from which the training images are cropped into $256\times256$ sub-images, with a step size of 128 on cropping. As a result, the total 800 training images produce about 340,000 training image samples, from which the image patches $\hat{X_i}$ (with a size of $64\times64$ each) are then further randomly cropped from these image samples. The patches are compressed by the PIL's JPEG encoder with a random quality factor $QF\in[1,60]$ to get the corresponding compressed patches $Y_i$. Such randomly generated training image pairs $\{\hat{X_i},Y_i\}$ can effectively help the model becoming more robust on handling artifacts removal under any quality factor setting imposed on the JPEG encoding stage.

To optimize our network, the Adam algorithm~\cite{Adam-2014} is used. Parameters are randomly initialized using a Gaussian distribution with a standard deviation of 0.01. The starting learning rate is set at 0.0001 and then decreased by a factor of 0.1 in every 20 epochs. The batch size is set to 256 for small-sized patches and 32 for large ones. For the other hyper-parameters of Adam in PyTorch, the default values are used. All the experiments are conducted on the GPU Tesla P100 with 16G bytes of memory.

\begin{table*}[!ht]
\renewcommand{\arraystretch}{1}
\centering
\setlength{\abovecaptionskip}{0pt}
\caption{
Performance comparisons of various methods based on the \textit{color} images from LIVE1 dataset~\cite{live1-2005} and BSDS500 dataset~\cite{bsds500-2011}, where the \textbf{boldface} denotes the best results, and the \underline{underline} indicates the second best.}
\label{tab:comparison}
\begin{tabular}{|c|c|c|c|c|c|c|c|}
  \hline
  \multirow{2}{*}{Dataset} & \multirow{2}{*}{Quality} & \multirow{2}{*}{Metric} & \multicolumn{5}{c|}{Methods} \\
  \cline{4-8}
  & & & JPEG & AR-CNN & EDSR & QCN & QGCN \\
  \hline
  \multirow{16}{*}{LIVE1}
  & \multirow{3}{*}{10}
  & PSNR & 25.60 & 26.28 & 27.44 & \underline{27.56} & \textbf{27.84} \\
  & & SSIM & 0.7454 & 0.7686 & 0.8008 & \underline{0.8035} & \textbf{0.8182} \\
  & & PSNR-B & 21.03 & 23.96 & 26.51 & \underline{26.73} & \textbf{27.01} \\
  \cline{2-8}
  & \multirow{3}{*}{20}
  & PSNR & 27.96 & 28.71 & 29.77 & \underline{29.94} & \textbf{30.16} \\
  & & SSIM & 0.8288 & 0.8473 & 0.8695 & \underline{0.8724} & \textbf{0.8831} \\
  & & PSNR-B & 23.20 & 26.99 & 28.31 & \underline{28.57} & \textbf{28.87} \\
  \cline{2-8}
  & \multirow{3}{*}{30}
  & PSNR & 29.25 & 29.96 & 31.06 & \underline{31.25} & \textbf{31.49} \\
  & & SSIM & 0.8642 & 0.8787 & 0.8979 & \underline{0.9006} & \textbf{0.9093} \\
  & & PSNR-B & 24.56 & 28.79 & 29.43 & \underline{29.60} & \textbf{29.94} \\
  \cline{2-8}
  & \multirow{3}{*}{40}
  & PSNR & 30.16 & 30.78 & 31.96 & \underline{32.17} & \textbf{32.39} \\
  & & SSIM & 0.8846 & 0.8960 & 0.9136 & \underline{0.9161} & \textbf{0.9236} \\
  & & PSNR-B & 25.61 & 29.97 & 30.29 & \underline{30.41} & \textbf{30.75} \\
  \cline{2-8}
  & \multirow{3}{*}{50}
  & PSNR & 30.91 & 31.39 & 32.66 & \underline{32.88} & \textbf{33.12} \\
  & & SSIM & 0.8990 & 0.9078 & 0.9244 & \underline{0.9267} & \textbf{0.9334} \\
  & & PSNR-B & 26.52 & 30.78 & 31.02 & \underline{31.09} & \textbf{31.43} \\
  \hline
  \multirow{16}{*}{BSDS500}
  & \multirow{3}{*}{10}
  & PSNR & 25.75 & 26.42 & 27.56 & \underline{27.66} & \textbf{27.90} \\
  & & SSIM & 0.7545 & 0.7767 & 0.8085 & \underline{0.8109} & \textbf{0.8160} \\
  & & PSNR-B & 20.73 & 23.63 & 26.22 & \underline{26.59} & \textbf{26.79} \\
  \cline{2-8}
  & \multirow{3}{*}{20}
  & PSNR & 28.11 & 28.84 & 29.84 & \underline{30.00} & \textbf{30.20} \\
  & & SSIM & 0.8402 & 0.8563 & 0.8772 & \underline{0.8798} & \textbf{0.8831} \\
  & & PSNR-B & 22.80 & 26.48 & 27.76 & \underline{28.10} & \textbf{28.32} \\
  \cline{2-8}
  &\multirow{3}{*}{30}
  & PSNR & 29.45 & 30.12 & 31.15 & \underline{31.32} & \textbf{31.51} \\
  & & SSIM & 0.8762 & 0.8876 & 0.9052 & \underline{0.9075} & \textbf{0.9101} \\
  & & PSNR-B & 24.11 & 28.20 & 28.73 & \underline{28.97} & \textbf{29.22} \\
  \cline{2-8}
  & \multirow{3}{*}{40}
  & PSNR & 30.40 & 30.96 & 32.07 & \underline{32.24} & \textbf{32.42} \\
  & & SSIM & 0.8966 & 0.9048 & 0.9209 & \underline{0.9230} & \textbf{0.9251} \\
  & & PSNR-B & 25.07 & 29.35 & 29.45 & \underline{29.61} & \textbf{29.85} \\
  \cline{2-8}
  & \multirow{3}{*}{50}
  & PSNR & 31.19 & 31.61 & 32.83 & \underline{33.00} & \textbf{33.18} \\
  & & SSIM & 0.9111 & 0.9167 & 0.9320 & \underline{0.9338} & \textbf{0.9358} \\
  & & PSNR-B & 25.98 & 30.32 & 30.21 & \underline{30.28} & \textbf{30.52} \\
  \hline
\end{tabular}
\end{table*}

\subsection{Performance Evaluation and Comparison}
Our proposed method is compared with two deep learning based methods on the performance of color image compression artifacts removal; i.e., AR-CNN~\cite{AR-CNN-2015} and EDSR~\cite{EDSR-2017}. The AR-CNN is the first method that exploit deep leaning to remove image artifacts from the decoded image. The EDSR is a single-image super-resolution method based on the residual network, and our proposed restoration network branch is motivated and developed from it. In fact, the EDSR network can be regarded as the ``baseline'' network that has no quantization tables as its supporting input compared to our proposed network. We can compare the performances of EDSR and that of our proposed QCN to demonstrate the benefits of using the quantization tables. To suit our objective, the EDSR network has been modified to adapt to image artifacts removal by adding a down-sampling layer in the beginning in order to produce an output image with the same size of the input image. For a fair comparison, all these methods use the same training settings as described in \secref{sec:implement}.

Our proposed network has two key novelties: (1) our network is trained by directly learning the quantization tables, and (2) our network has added a global network branch to assist the main restoration branch. To evaluate the performance of each novelty, two variants of network are constructed. The network with quantization table learning is denoted as the quantization-table convolutional network (QCN), and the QCN with global network branch is denoted as QGCN. These two networks are trained under the same settings to evaluate the effectiveness of our proposed method.

Two datasets are tested to evaluate both the quantitative and qualitative performance, including LIVE1 dataset~\cite{live1-2005} (29 images) and BSDS500 dataset~\cite{bsds500-2011} (200 images in the test set). Three image quality assessment criteria, PSNR, structural similarity (SSIM)~\cite{SSIM-2004}, and PSNR-B~\cite{PSNR-B-2011}, are chosen to conduct quantitative evaluation. Note that the PSNR-B is specially designed to evaluate the \textit{blocking artifacts} presented in image; it is quite effective on the evaluation of decoded image quality. The evaluation results performed on color images of LIVE1 and BSDS500 dataset are shown in \tabref{tab:comparison}; from which, one can see that QCN is better than EDSR, and QGCN is better than QCN, in all three image quality assessments. This shows that the quantization table based learning and the global optimization network are both effective for compression artifacts removal. In what follows, we shall give two examples to demonstrate that our developed QGCN is able to handle both \textit{global} structure and \textit{local} structure quite well on image artifacts removal; these are demonstrated in \figref{fig:compare1} and \figref{fig:compare2}, respectively.

\figref{fig:compare1} presents a set of image artifacts removal results, obtained from different methods. Through this example, it is our goal to highlight that our method is able to handle \textit{global} structure quite well. The image in \sfigref{fig:compare1} (a) is compressed by using the JPEG with the quality factor of 10, and the decoded image is shown in \sfigref{fig:compare1} (b). One of the primary image artifacts caused by setting a large degree of image compression (i.e., using a lower value of the quality factor like 10 in this case) is the well-known \textit{blocking} artifacts; for example, refer to the sky area in \sfigref{fig:compare1} (b) that has distinctly visible contours. Equally serious, the original color is also changed due to the distortion incurred by the compression on the chrominance components. In \sfigref{fig:compare1} (b), the overall color tone turns out to be a little reddish and darker. In \sfigref{fig:compare1} (c) and (d), the results of the AR-CNN and EDSR have shown distinct contours and false colors in the water areas. On the other hand, our QCN and QGCN in \sfigref{fig:compare1} (e) and (f) are able to effectively remove image artifacts. To be more specific, without global optimization, the color presented on the QCN is still a bit reddish. With global optimization, our QGCN is able to restore the color fairly close to the original (ground-truth) image. This example illustrates that our method with global optimization has advantage in removing blocking artifacts in a large smooth or flat area. Furthermore, the global optimization is able to restore the color distribution of the compressed image to a more natural and visually pleasant presentation.

\figref{fig:compare2} illustrates another comparison results of image artifacts removal with emphasis on demonstrating the proposed method is able to handle \textit{local} structure quite well. The test image \textit{Cemetry} is encoded using the same quality factor of 10 imposed on the JPEG compression stage. Two small image patches are selected and zoomed in order to show the image details. The image patch of English words shows distinct ringing artifacts around the white letters. The image patch of the gate also shows ringing artifacts, plus blocking artifacts on the grass background. In \sfigref{fig:compare2} (f), our QGCN method effectively removes these artifacts and is superior to others. The local structures of the words and gate are restored correctly. This example shows that our method is able to deal with local structure of the compressed image very well.

Most existing compression artifacts removal methods train and evaluate their models base on grayscale images (i.e., Y channel of YCbCr color space) only. To further demonstrate the effectiveness of our method, experiments on grayscale images are also conducted to compare with the existing methods. The methods to be compared are AR-CNN~\cite{AR-CNN-2015}, DnCNN~\cite{zhang2017DnCNN}, CAS-CNN~\cite{cavigelli2017cas}, DMCNN~\cite{zhang2018dmcnn}, IDCN~\cite{zheng2019IDCN} and RNAN~\cite{zhang2019RNAN}. However, most of these methods train there networks on specific values of quality factor. For a fair comparison, we improved their published codes and retrained these competitive methods under the same settings. The methods of AR-CNN, DnCNN, CAS-CNN, IDCN and our QGCN are all trained with one model over the same quality factors, ranging from 1 to 60. However, it is important to point out here that the method of DMCNN is unable to be trained for a range of quality factors, but on specific values of quality factor; this is due to the structure limitation of their designed network. To remedy this limitation, five specific quality factors, 10, 20, 30, 40, and 50, are individually trained for it and tested likewise. The evaluation results are documented in \tabref{tab:compareY}. The best results are highlighted in boldface and underlined for the second best. From this table, one can see that our proposed QGCN mostly outperforms DMCNN and performs significantly better than other methods for all quality factors.

\begin{figure}[t]
  \centering
  \includegraphics[width=\linewidth]{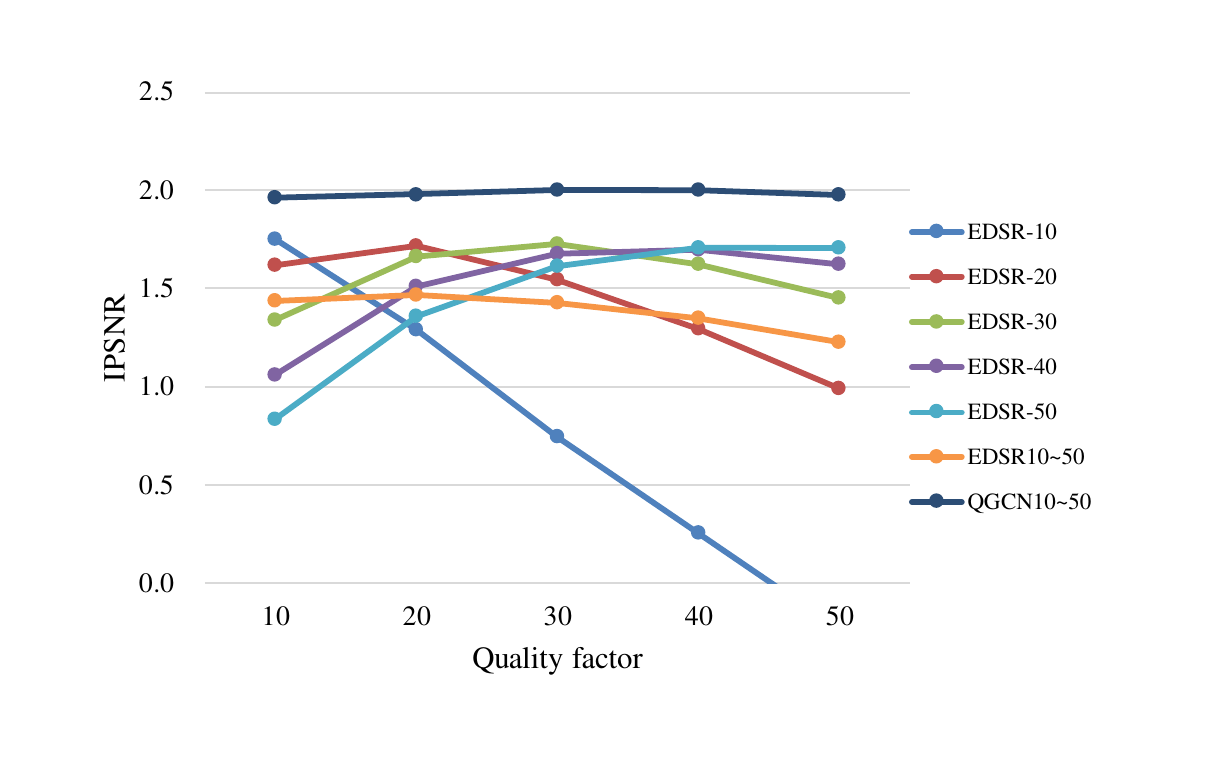}\\
  \caption{A study of the generalization ability or robustness of our proposed method. The training of EDSR~\cite{EDSR-2017} model requires a specific quality factor determined in advance, and the performance will be dropped quickly for mismatched quality-factor cases. With wider range of quality factors being trained, a stable but lower performance is achieved as shown by EDSR10-50. On the other hand, our QGCN model not only yields a stable performance across all quality factors but gets the best performance.}
  \label{fig:generalize}
\end{figure}

\begin{table*}[t]
\renewcommand{\arraystretch}{1}
\centering
\setlength{\abovecaptionskip}{0pt}
\caption{Performance comparisons of various methods based on the \textit{grayscale} images from LIVE1 dataset~\cite{live1-2005} and BSDS500 dataset~\cite{bsds500-2011}, where the \textbf{boldface} denotes the best results, and the \underline{underline} indicates the second best.}
\label{tab:compareY}
\begin{tabular}{|c|c|c|c|c|c|c|c|c|c|c|}
  \hline
  \multirow{2}{*}{Dataset} & \multirow{2}{*}{Quality} & \multirow{2}{*}{Metric} & \multicolumn{8}{c|}{Methods} \\
  \cline{4-11}
  & & & JPEG & AR-CNN & DnCNN-3 & CAS-CNN & DMCNN & IDCN & RNAN & QGCN \\
  \hline
  \multirow{16}{*}{LIVE1}
  & \multirow{3}{*}{10}
  & PSNR & 27.77 & 28.60 & 29.19 & 29.43 & \textbf{29.74} & 29.62 & 29.39 & \textbf{29.74} \\
  & & SSIM & 0.7905 & 0.8148 & 0.8123 & 0.8339 & \underline{0.8395} & 0.8375 & 0.8328 & \textbf{0.8399} \\
  & & PSNR-B & 25.33 & 27.58 & 28.90 & 28.98 & \textbf{29.43} & 29.24 & 29.01 & \underline{29.40} \\
  \cline{2-11}
  & \multirow{3}{*}{20}
  & PSNR & 30.07 & 31.10 & 31.59 & 31.82 & \underline{32.08} & 31.94 & 31.78 & \textbf{32.11} \\
  & & SSIM & 0.8683 & 0.8856 & 0.8802 & 0.8981 & \underline{0.9010} & 0.8992 & 0.8973 & \textbf{0.9017} \\
  & & PSNR-B & 27.57 & 30.35 & 31.07 & 31.17 & \underline{31.50} & 31.35 & 31.17 & \textbf{31.56} \\
  \cline{2-11}
  & \multirow{3}{*}{30}
  & PSNR & 31.41 & 32.46 & 32.98 & 33.20 & 33.40 & \underline{33.41} & 33.18 & \textbf{33.53} \\
  & & SSIM & 0.9000 & 0.9120 & 0.9090 & 0.9236 & \underline{0.9253} & 0.9252 & 0.9232 & \textbf{0.9265} \\
  & & PSNR-B & 28.92 & 31.89 & 32.34 & 32.42 & \underline{32.78} & 32.07 & 32.41 & \textbf{32.81} \\
  \cline{2-11}
  & \multirow{3}{*}{40}
  & PSNR & 32.35 & 33.37 & 33.96 & 34.17 & \underline{34.39} & 34.37 & 34.15 & \textbf{34.50} \\
  & & SSIM & 0.9173 & 0.9257 & 0.9247 & 0.9372 & \underline{0.9387} & 0.9384 & 0.9370 & \textbf{0.9397} \\
  & & PSNR-B & 29.96 & 32.95 & 33.28 & 33.33 & \textbf{33.74} & 33.57 & 33.31 & \underline{33.71} \\
  \cline{2-11}
  & \multirow{3}{*}{50}
  & PSNR & 33.16 & 34.09 & 34.77 & 34.94 & 35.21 & \underline{35.22} & 34.96 & \textbf{35.32} \\
  & & SSIM & 0.9295 & 0.9348 & 0.9356 & 0.9462 & \underline{0.9479} & 0.9478 & 0.9464 & \textbf{0.9488} \\
  & & PSNR-B & 30.86 & 33.74 & 34.06 & 34.09 & \textbf{34.53} & 34.33 & 34.05 & \underline{34.46} \\
  \hline
  \multirow{16}{*}{BSDS500}
  & \multirow{3}{*}{10}
  & PSNR & 27.80 & 28.62 & 29.21 & 29.44 & \textbf{29.67} & 29.55 & 29.15 & \underline{29.65} \\
  & & SSIM & 0.7875 & 0.8121 & 0.8090 & 0.8314 & \underline{0.8363} & 0.8341 & 0.8272 & \textbf{0.8365} \\
  & & PSNR-B & 25.10 & 27.38 & 28.80 & 28.87 & \textbf{29.28} & 29.11 & 28.62 & \underline{29.21} \\
  \cline{2-11}
  & \multirow{3}{*}{20}
  & PSNR & 30.05 & 31.05 & 31.53 & 31.77 & \underline{31.99} & 31.83 & 31.47 & \textbf{32.01} \\
  & & SSIM & 0.8671 & 0.8840 & 0.8775 & 0.8964 & \underline{0.8989} & 0.8969 & 0.8929 & \textbf{0.8990} \\
  & & PSNR-B & 27.22 & 30.01 & 30.79 & 30.83 & \textbf{31.19} & 31.05 & 30.76 & \textbf{31.19} \\
  \cline{2-11}
  & \multirow{3}{*}{30}
  & PSNR & 31.37 & 32.39 & 32.90 & 33.13 & 33.24 & \underline{33.27} & 32.85 & \textbf{33.36} \\
  & & SSIM & 0.8994 & 0.9110 & 0.9069 & 0.9226 & \underline{0.9235} & \underline{0.9235} & 0.9196 & \textbf{0.9247} \\
  & & PSNR-B & 28.53 & 31.52 & 31.97 & 31.93 & \textbf{32.34} & 32.23 & 32.02 & \underline{32.33} \\
  \cline{2-11}
  & \multirow{3}{*}{40}
  & PSNR & 32.30 & 33.29 & 33.85 & 34.07 & \underline{34.20} & \underline{34.20} & 33.82 & \textbf{34.32} \\
  & & SSIM & 0.9171 & 0.9250 & 0.9230 & 0.9365 & \underline{0.9374} & 0.9371 & 0.9341 & \textbf{0.9383} \\
  & & PSNR-B & 29.49 & 32.53 & 32.80 & 32.71 & \textbf{33.18} & 32.99 & 32.92 & \underline{33.12} \\
  \cline{2-11}
  & \multirow{3}{*}{50}
  & PSNR & 33.10 & 34.03 & 34.67 & 34.88 & 35.02 & \underline{35.06} & 34.67 & \textbf{35.14} \\
  & & SSIM & 0.9296 & 0.9347 & 0.9346 & 0.9462 & \underline{0.9472} & \underline{0.9472} & 0.9444 & \textbf{0.9480} \\
  & & PSNR-B & 30.38 & 33.39 & 33.60 & 33.46 & \textbf{33.96} & 33.74 & 33.76 & \underline{33.85} \\
  \hline
\end{tabular}
\end{table*}

\begin{figure*}[!ht]
  \centering
  \includegraphics[width=1\textwidth]{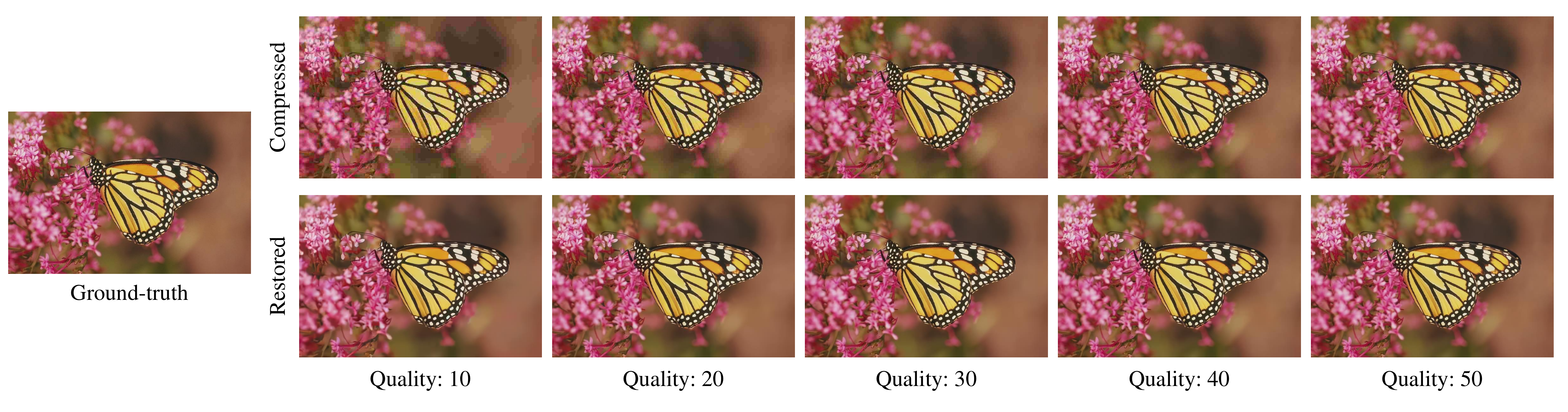}\\
  \caption{The artifacts removal results of multiple compression qualities of the butterfly image under the quality factors from 10 to 50. Note that all the results are produced by a single learned model.}
  \label{fig:multires}
\end{figure*}

\begin{figure}[t]
  \centering
  \includegraphics[width=\linewidth]{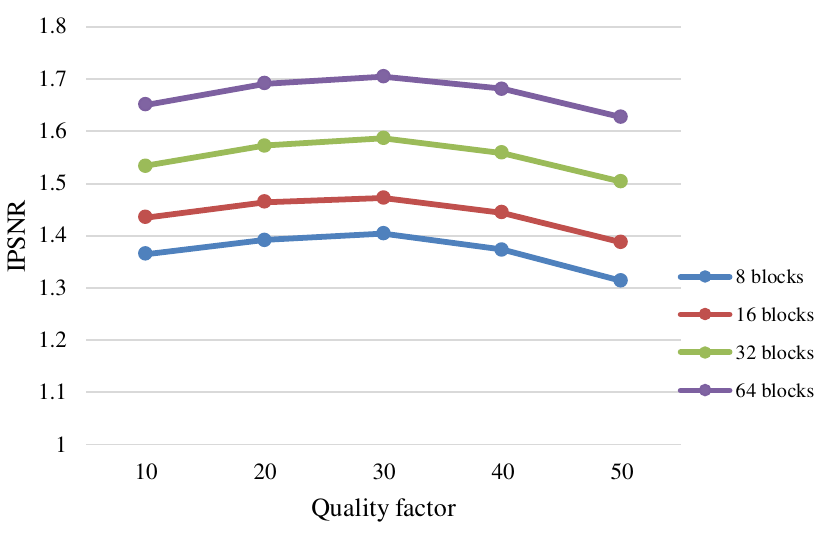}\\
  \caption{Comparison on different sizes of models. They show stable performance across a wide range of quality factors (from 10 to 50) in each case, regardless the size of the model.}
  \label{fig:modelsize}
\end{figure}

\subsection{Performance Analysis}
\subsubsection{Generalization ability}
The proposed method has an outstanding generalization ability. It can learn a single model to effectively remove image artifacts from the de-compressed images that were encoded by using the JPEG algorithm with any chosen quality factor. As a comparison, the EDSR network is trained on five specific quality factors, ranging from 10 to 50 (with a step size of 10), and tested on all these five quality factors. To see how much PSNR improvements \textit{before} and \textit{after} image artifacts removal, the metric named \textit{incremental} PSNR (IPSNR)~\cite{ multi-test2016} is used here, which is the difference of two PSNR measurements---i.e., the degraded (i.e., ``before'') image and the restored (i.e., ``after'') image, with respect to the original (ground-truth) image, respectively. \figref{fig:generalize} shows the IPSNR comparisons of our \textit{arbitray}-quality model and other \textit{single}-quality models. From which, one can see that the single-quality models only perform well on the assumed quality factor, and drop quickly when the test images were compressed under other quality factors. On the other hand, our QGCN model achieves a stable performance across all quality factors (from 10 to 50) and outperforms the other methods, even trained and tested on the same quality factor. This clearly demonstrates outstanding robustness and excellent generalization ability of our proposed method.

\figref{fig:multires} shows a set of compressed images \textit{Butterfly} with the quality factors ranging from 10 to 50 in the first row, and the corresponding restored images using our proposed arbitrary-quality single model approach are shown in the second row correspondingly. One can see that our method can handle all cases and achieve impressive restoration results consistently, regardless which quality factor encountered.

\begin{figure}[t]
  \centering
  \includegraphics[width=\linewidth]{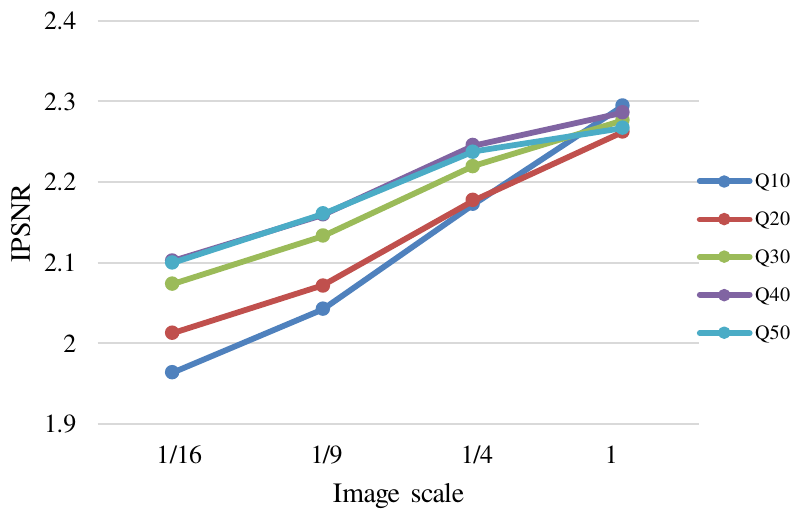}\\
  \caption{Comparison on different resolution images. Our method gets a better restoration performance on higher resolution images, which benefits from the global network branch.}
  \label{fig:highres}
\end{figure}

There may be a concern that whether the generalization ability is related with the large size of the model but not the quantization map.
To address this concern, we set up a comparison of models with different numbers of residual blocks. Four models with 8, 16, 32, and 64 residual blocks, respectively, are trained under the same setting and conditions. \figref{fig:modelsize} shows the evaluation results of these models. From the figure one can see that all the models are able to achieve a stable performance across quality factors from 10 to 50. This demonstrates the robustness and generalization ability of our method, which has nothing to do with the size of the model.

\subsubsection{High-resolution images}
The proposed QGCN network with global branch learns image features from the entire image to effectively assist the restoration network branch and further improve the image artifacts removal results. To evaluate the effectiveness contributed from the global network branch, our QGCN method is tested on images with different resolutions.
The high-resolution images from the DIV2K dataset~\cite{DIV2K-2017} are used as our test images, and resized into three different resolutions, i.e., $1/4$, $1/9$, and $1/16$ of the original area size. Five compression quality factors (from 10 to 50) are applied to these test images, followed by exploiting our QGCN model to remove image artifacts for each decoded image. The resulted IPSNR is shown in \figref{fig:highres}; from which, one can see that the IPSNR will be increased, if the image resolution is increased. This result shows that our model performs better for high-resolution image, which is benefitted from our proposed global network branch.

\subsubsection{The size of training patches}
\begin{figure}[t]
  \centering
  \includegraphics[width=\linewidth]{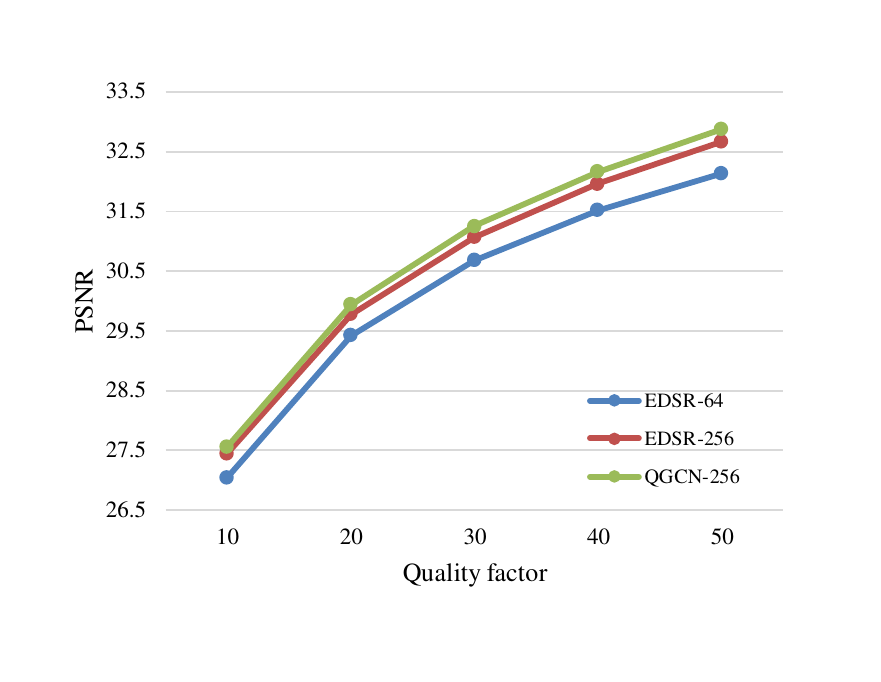}\\
  \caption{Comparison on different sizes of training patches. It shows that larger training patches can improve the performance of compression artifacts removal. Our method achieves better performance under the same patch size ($256\times256$).}
  \label{fig:patchsize}
\end{figure}

Our model is first trained on image patches with a size of $64\times64$, then fine-tuned on patches with a size of $256\times256$. To evaluate the effect of training patch size, we compare our model with an EDSR model trained on the image patches with the same size. The PSNR results is shown in \figref{fig:patchsize}. The PSNR values of EDSR-256 is better than that of EDSR-64, which means that larger-sized training images can yield better performance. Our QGCN-256 model is better than EDSR-256 in PSNR values, which shows the effectiveness of our method.

\section{Conclusion}
\label{sec:conclude}
In this paper, a \textit{single}-model convolutional neural network for conducting image artifacts removal of the JPEG-decoded images is proposed. Apart from the existing state-of-the-art CNN-based approaches, our proposed network is able to deal with any JPEG-compressed image with wide (even up to full) range of quality factors imposed during the compression stage. To tackle this challenge, our \emph{restoration branch} sub-network is trained with the use of \textit{quantization tables}; this prior information covers all possible quality factors made available in the JPEG. The established single-model CNN makes our image artifacts removal task much more efficient, robust, and practical. With the goal of pursuing global optimization, our \emph{global branch} sub-network augments the performance of the above-mentioned restoration branch sub-network and further improves the image quality at the final restored image. It has been shown that the proposed global branch is especially beneficial to high-resolution image cases; this is in line with the current technology trend on the evolution and uses of high-definition image processing across a plethora of image-based applications.


\ifCLASSOPTIONcaptionsoff
  \newpage
\fi




\bibliographystyle{IEEEtran}
\bibliography{artifacts}




\end{document}